\documentclass[12pt]{article}
\usepackage[margin=2.5cm]{geometry}
\usepackage{amsmath}
\usepackage{amssymb}
\usepackage[utf8]{inputenc}
\usepackage{bbold}
\usepackage{xcolor}
\usepackage[makeroom]{cancel}
\usepackage{mathtools}
\usepackage{bbm}
\usepackage[colorlinks, allcolors=cyan]{hyperref}

\usepackage[backend=biber, style=phys, sorting=none, citestyle=numeric-comp, url=false, eprint=false]{biblatex}
\appto{\bibsetup}{\sloppy}

\bibliography{bibliography}
\title{On the Unification of Conformal and Fuzzy Gravities with $SO(10)$ GUT}

\begin{document}
\author{Stelios Stefas$^1$, George Zoupanos$^{1,2,3,4}$}\date{}

\maketitle
\begin{center}
\itshape$^1$ Physics Department, National Technical University of Athens, Zografou Campus, 157 80, Zografou, Greece\\
\itshape$^2$ Max-Planck Institut f\'ur Physik, Boltzmannstr. 8, 85 748 Garching/Munich, Germany\\
\itshape$^3$ Universit\"at Hamburg, Luruper Chaussee 149, 22 761 Hamburg, Germany\\
\itshape$^4$  Deutsches Elektronen-Synchrotron DESY, Notkestra{\ss}e 85, 22 607, Hamburg, Germany
\end{center}

\begin{center}
\emph{E-mails: \href{mailto:dstefas@mail.ntua.gr}{dstefas@mail.ntua.gr}, \href{mailto:george.Zoupanos@cern.ch}{george.zoupanos@cern.ch}}
\end{center}

\begin{abstract}
Within the gauge-theoretic approach of gravity, the gauging of an enlarged symmetry of the tangent space in four dimensions allows gravity to be unified with internal interactions. We study the unification of the Conformal and Noncommutative (Fuzzy) Gravities with Internal Interactions based on the $SO(10)$ GUT.
\end{abstract}

\section{Introduction}

It is well established that the Standard Model (SM) of particle physics is elegantly described by gauge theories. It has also been long recognized that General Relativity (GR) can be reformulated as a gauge theory \cite{utiyama, kibble1961, macdowell, Ivanov:1980tw, Ivanov:1981wm, stellewest, Kibble:1985sn}. In our recent works \cite{Manolakos:2023hif, Konitopoulos:2023wst, Roumelioti:2024lvn, roumelioti2407}, we proposed the exploration of the possibility of unifying gravity with the other fundamental interactions within a gauge-theoretic framework (see also \cite{Percacci:1984ai, Percacci_1991, Nesti_2008, Nesti_2010, Chamseddine2010, Chamseddine2016, Krasnov:2017epi, noncomtomos}). The key observation is that, while the dimension of the tangent space is typically taken to match that of the underlying spacetime manifold, the tangent group of a $d$-dimensional manifold need not be restricted to $SO(d)$ \cite{Weinberg:1984ke}. This opens the road towards considering higher-dimensional tangent groups in a four-dimensional spacetime, potentially leading to a natural unification of gravity and internal gauge interactions by gauging such extended tangent groups.

A notable advantage of this framework is that it allows the tools developed for higher-dimensional theories with extra spacetime dimensions—such as those employed in the Coset Space Dimensional Reduction (CSDR) program \cite{forgacs, MANTON1981502, CHAPLINE1982461,LUST1985309, Kubyshin:1989vd, KAPETANAKIS19924, Manousselis_2004, Chatzistavrakidis:2009mh, Irges:2011de, Manolakos:2020cco}—to be directly applied in four dimensions, since the tangent group remains the same. In particular, this approach inherits the familiar constraints needed to construct realistic chiral gauge theories describing the internal symmetries. For example, to obtain a chiral gauge theory in four dimensions from a higher-dimensional one, the Weyl condition must be imposed. Moreover, in order to reduce the number of fermion families, in principle, the Majorana condition can also be imposed in certain dimensions \cite{CHAPLINE1982461, KAPETANAKIS19924}.

Building on these ideas, a recent proposal has realized a unification of Conformal Gravity (CG) with internal gauge interactions \cite{Roumelioti:2024lvn}. This framework has been further extended to include four-dimensional gravity formulated on a covariant noncommutative (fuzzy) space, leading to a unification of Fuzzy Gravity (FG) with internal interactions \cite{roumelioti2407}. In what follows, we summarize the essential features of these constructions.

\section{Conformal Gravity}

To begin with, we recall that Einstein Gravity (EG) can be successfully viewed as a gauge theory of the Poincar\'e group \cite{kibble1961}. Greater conceptual clarity and elegance, however, is achieved by gauging the de Sitter (dS) and anti-de Sitter (AdS) groups, $SO(1,4)$ and $SO(2,3)$, respectively. Both groups possess 10 generators, just like the Poincar\'e group, and can be spontaneously broken to the Lorentz group $SO(1,3)$ via the introduction of a scalar field \cite{stellewest, Kibble:1985sn, Roumelioti:2024lvn, manolakosphd}. Moreover, the Poincar\'e, dS, and AdS groups are all subgroups of the conformal group $SO(2,4)$, which has 15 generators. In \cite{Kaku:1978nz}, the gauge-theoretic formulation of gravity was extended to $SO(2,4)$, leading to CG. The breaking of CG to EG or to Weyl’s scale-invariant gravity was originally achieved by imposing algebraic constraints on the gauge fields \cite{Kaku:1978nz}. In contrast, \cite{Roumelioti:2024lvn} achieved this breaking dynamically for the first time, by introducing a scalar field into the action and employing the Lagrange multiplier method. 

The gauge theory for CG, as it was mentioned above, is based on $SO(2,4)$ as the gauge group, which is isomorphic to $SU(4)$ as well as $SO(6)$ (In what follows we will work in the Euclidean signature for convenience). In order to perform its spontaneous symmetry breaking (SSB), one can follow two distinct routes.

The first route, leading to EG, would be to introduce a scalar field in the vector representation (rep) of $SO(6)$, namely the $\mathbf{6}$, which acquires a vacuum expectation value (vev) along its $\langle \mathbf{1} \rangle$ component. This is seen from the branching of the $\mathbf{6}$ of $SO(6)$ under its maximal subgroup $SO(5)$ \cite{Slansky:1981yr, Feger_2020, Roumelioti:2024lvn}: 
\begin{equation}
\label{SO6toSO5}
\begin{aligned}
SO(6) &\supset SO(5) ,\\
\mathbf{6} &= \mathbf{1} + \mathbf{5}\, .
\end{aligned}
\end{equation}
The unbroken $SO(5)$, isomorphic to $SO(2,3)$, may then break further to $SO(1,3)$ when a scalar in the $\mathbf{5}$ obtains a vev in the $\langle \mathbf{1},\mathbf{1} \rangle$ component according to:
\begin{equation}
\label{SO5toSU2SU2}
\begin{aligned}
SO(5) &\supset SU(2) \times SU(2) ,\\
\mathbf{5} &= (\mathbf{1}, \mathbf{1}) + (\mathbf{2}, \mathbf{2})\, ,
\end{aligned}
\end{equation}
where the algebra of $SU(2) \times SU(2)$ is isomorphic to those of $SO(4)$ and $SO(1,3)$. Consequently, two scalar fields in the vector rep $\mathbf{6}$ of $SO(2,4)$ are required to realize the full breaking $SO(2,4) \to SO(1,3)$ (see \cite{Roumelioti:2024lvn} for details).

Alternatively, the second route of breaking $SO(2,4) \to SO(1,3)$ can be achieved directly in a single step by introducing a scalar in the second-rank antisymmetric rep $\mathbf{15}$ of $SO(6)\sim SO(2,4)$. Depending on the chosen vacuum, this procedure yields either EG or Weyl Gravity (WG), as will be discussed later in this section.

More specifically, the gauge group $SO(2,4)$ has 15 generators, which, written in four-dimensional notation, can be expressed as six Lorentz transformations $M_{ab}$, four translations $P_a$, four conformal boosts $K_a$, and one dilatation $D$. A gauge connection $A_\mu$ is introduced, which, being an $SO(2,4)$-valued one-form, can be expanded on the above generators as
\begin{equation}
A_\mu=\frac{1}{2}\omega_\mu{}^{ab} M_{ab} + e_\mu{}^a P_a + b_\mu{}^a K_a + \tilde{a}_\mu D ,
\end{equation}
where each generator is associated with a gauge field. In particular, the vierbein $e_\mu{}^a$ and the spin connection $\omega_\mu{}^{ab}$ are identified with the gauge fields of translations and Lorentz transformations respectively, while there is also one special conformal gauge field $b_\mu{}^a$, as well as a dilatation gauge field $\tilde{a}_\mu$. Accordingly, the field strength tensor takes the form
\begin{equation}
\label{fieldstrengthconformal}
F_{\mu\nu} = \frac{1}{2}R_{\mu\nu}{}^{ab}M_{ab} + \tilde{R}_{\mu\nu}{}^{a}P_a + R_{\mu\nu}{}^{a}K_a + R_{\mu\nu}D ,
\end{equation}
in which the usual 4D curvature and torsion tensors are included\footnote{The explicit expressions of the components of the field strength tensor can be found in \cite{Roumelioti:2024lvn}.}. 

A parity-conserving action is considered, quadratic in the field strength \eqref{fieldstrengthconformal}, in which a scalar $\phi$ in the second rank antisymmetric rep $\mathbf{15}$ of $SO(6)\sim SO(2,4)$ as well as a dimensionful parameter $m$ are introduced:
\begin{equation}
S_{SO(2,4)} = a_{CG} \int d^4x \left[\operatorname{tr}\,\epsilon^{\mu\nu\rho\sigma} m \phi F_{\mu\nu}F_{\rho\sigma} + \left(\phi^2 - m^{-2}\mathbb{I}_4\right)\right],
\end{equation}
with the trace defined as $\mathrm{tr} \to \epsilon_{abcd}[\text{Generators}]^{abcd}$. Furthermore, the scalar $\phi$, being an element of the algebra, can be expanded on the generators as
\begin{equation}
\phi = \phi^{ab}M_{ab} + \tilde{\phi}^a P_a + \phi^a K_a + \tilde{\phi}D .
\end{equation}
Following \cite{Li:1973mq}, we work in the gauge where $\phi$ is diagonal, $\mathrm{diag}(1,1,-1,-1)$, and purely aligned with the dilatation generator $D$:
\begin{equation}
\phi = \tilde{\phi}D \xrightarrow{\phi^2=m^{-2}\mathbb{I}_4} \phi = -2m^{-1}D .
\end{equation}
In this particular gauge, an SSB is induced and the action reduces to
\begin{equation}
S = -2a_{CG} \int d^4x\, \operatorname{tr}\,\epsilon^{\mu\nu\rho\sigma}F_{\mu\nu}F_{\rho\sigma}D ,
\end{equation}
where the gauge fields $e, b, \tilde{a}$ rescale as $me,\ mb,\ m\tilde{a}$, respectively. After a straightforward calculation, using the expansion of $F_{\mu\nu}$ and the commutation and anticommutation relations of the generators, one finds \cite{Roumelioti:2024lvn}:
\begin{equation}
S_{SO(1,3)} = \frac{a_{CG}}{4}\int d^4x\, \epsilon^{\mu\nu\rho\sigma}\epsilon_{abcd}R_{\mu\nu}{}^{ab}R_{\rho\sigma}{}^{cd},
\end{equation}
which is manifestly only Lorentz-invariant. A closer look in the broken action reveals that the field $\tilde{a}_\mu$ is absent from the expression in which allows us to set $\tilde{a}_\mu=0$, leading to simplified expressions for the $P$- and $K$-sector field strength components:
\begin{equation}
\begin{aligned}
\tilde{R}_{\mu \nu}{ }^a & =m T_{\mu \nu}^{(0) a}(e)-2 m^2 \tilde{a}_{[\mu} e_{\nu]}{ }^a \longrightarrow m T_{\mu \nu}^{(0) a}(e), \\
R_{\mu \nu}{ }^a & =m T_{\mu \nu}^{(0) a}(b)+2 m^2 \tilde{a}_{[\mu} b_{\nu]}{ }^a \longrightarrow m T_{\mu \nu}^{(0) a}(b) .
\end{aligned}
\end{equation}
where $T_{\mu\nu}^{(0)a}$ is the torsion tensor. Furthermore, the absence of these components in the action allows us to impose $\tilde{R}_{\mu\nu}{}^{a}=R_{\mu\nu}{}^{a}=0$, thus yielding a torsion-free theory. Since $R_{\mu\nu}$ also does not appear in the action, we set $R_{\mu\nu}=0$, which imposes the relation
\begin{equation}
\label{e-b-relation}
e_\mu{}^{a}b_{\nu a} - e_\nu{}^{a}b_{\mu a} = 0 .
\end{equation}
This motivates us to consider solutions relating $e$ and $b$. Two notable cases are:

\textbf{A.} For $\boxed{b_\mu{}^{a}=a\,e_\mu{}^{a}}$, one obtains the Einstein–Hilbert action with a cosmological constant:
\begin{equation}
\begin{aligned}
S_{\mathrm{SO}(1,3)}=\frac{a_{C G}}{4} \int d^4 x \epsilon^{\mu \nu \rho \sigma} \epsilon_{a b c d}\left[R_{\mu \nu}^{(0) a b} R_{\rho \sigma}^{(0) c d}\right. & -16 m^2 a R_{\mu \nu}^{(0) a b} e_\rho{ }^c e_\sigma^d \\
& \left.+64 m^4 a^2 e_\mu{ }^a e_\nu^b e_\rho{ }^c e_\sigma^d\right]\, .
\end{aligned}
\end{equation}

\textbf{B.} For $\boxed{b_\mu{}^{a}= -\frac{1}{4}\left(R_\mu{}^{a}-\frac{1}{6}Re_\mu{}^{a}\right)}$, one obtains the Weyl action:
\begin{equation}
S = \frac{a_{CG}}{4}\int d^4x\, \epsilon^{\mu\nu\rho\sigma}\epsilon_{abcd}C_{\mu\nu}{}^{ab}C_{\rho\sigma}{}^{cd}
= 2a_{CG}\int d^4x\left(R_{\mu\nu}R^{\nu\mu}-\frac{1}{3}R^2\right),
\end{equation}
where $C_{\mu\nu}{}^{ab}$ is the Weyl conformal tensor.
\section{Noncommutative Gauge Theory of 4D Gravity -- Fuzzy Gravity}

\subsection{The Background Space}\

Before constructing the gauge theory of Fuzzy Gravity, we will first have to specify the background space on which it is going to be defined. Building on Snyder's original proposal \cite{Snyder:1946qz} and its extensions \cite{yang1947, Heckman_2015, Manolakos_paper1, Manolakos_paper2, Manolakos:2022universe}, one considers the group $SO(1,5)$ and identifies the four-dimensional spacetime coordinates with elements of its Lie algebra.

Specifically, the $SO(1,5)$ generators satisfy
\begin{equation}
\left[J_{mn}, J_{rs}\right] = i \left( \eta_{mr} J_{ns} + \eta_{ns} J_{mr} - \eta_{nr} J_{ms} - \eta_{ms} J_{nr} \right),
\end{equation}
where $m,n,r,s = 0,\dots,5$ and $\eta_{mn} = \mathrm{diag}(-1,1,1,1,1,1)$. Decomposing $SO(1,5)$ down to $SO(1,3)$ via the chains $SO(1,5) \supset SO(1,4) \supset SO(1,3)$ yields the algebra:
\begin{equation}
\begin{gathered}
\left[J_{ij},J_{kl}\right] = i \left(\eta_{ik}J_{jl} + \eta_{jl}J_{ik} - \eta_{jk}J_{il} - \eta_{il}J_{jk}\right), \quad
\left[J_{ij}, J_{k5}\right] = i \left(\eta_{ik}J_{j5} - \eta_{jk}J_{i5}\right), \\
\left[J_{i5}, J_{j5}\right] = i J_{ij}, \quad 
\left[J_{ij}, J_{k4}\right] = i \left(\eta_{ik}J_{j4} - \eta_{jk}J_{i4}\right), \quad 
\left[J_{i4}, J_{j4}\right] = i J_{ij}, \\
\left[J_{i4}, J_{j5}\right] = i \eta_{ij} J_{45}, \quad 
\left[J_{ij}, J_{45}\right] = 0, \quad 
\left[J_{i4}, J_{45}\right] = -i J_{i5}, \quad 
\left[J_{i5}, J_{45}\right] = i J_{i4}.
\end{gathered}
\end{equation}
Consequently, the generators can be converted to physical quantities, by identifying them with the noncommutativity tensor, the coordinates and momenta via
\begin{equation}
\Theta_{ij} = \hbar J_{ij}, \quad X_i = \lambda J_{i5}, \quad P_i = \frac{\hbar}{\lambda} J_{i4}, \quad h = J_{45},
\end{equation}
respectively, where $\lambda$ is a natural length scale. This leads to the following commutation relations:
\begin{equation}
\begin{gathered}
[\Theta_{ij}, \Theta_{kl}] = i\hbar (\eta_{ik}\Theta_{jl} + \eta_{jl}\Theta_{ik} - \eta_{jk}\Theta_{il} - \eta_{il}\Theta_{jk}), \\
[\Theta_{ij}, X_k] = i\hbar (\eta_{ik} X_j - \eta_{jk} X_i), \quad [\Theta_{ij}, P_k] = i\hbar (\eta_{ik} P_j - \eta_{jk} P_i), \\
[X_i, X_j] = \frac{i\lambda^2}{\hbar} \Theta_{ij}, \quad [P_i, P_j] = \frac{i\hbar}{\lambda^2} \Theta_{ij}, \quad [X_i, P_j] = i \hbar \eta_{ij} h, \\
[\Theta_{ij}, h] = 0, \quad [X_i, h] = \frac{i\lambda^2}{\hbar} P_i, \quad [P_i, h] = -\frac{i\hbar}{\lambda^2} X_i.
\end{gathered}
\end{equation}

From these relations, one observes that both spacetime coordinates and momenta are noncommutative, implying quantization of spacetime and momentum space. Moreover, the commutator between coordinates and momenta naturally reproduces a Heisenberg-type commutation relation.

\subsection{Gauge Group and Representation}\

To formulate a gauge theory of gravity on this background space, one must first choose the appropriate gauge group. The natural choice would be the group that describes the symmetries of the theory, i.e. the isometry group of $dS_4$, $SO(1,4)$. However, in noncommutative gauge theories, anticommutators of gauge generators appear inevitably. Since the anticommutators of $SO(1,4)$ generators do not necessarily close within the original algebra, a suitable rep must be chosen and the gauge group has to be extended in order to include the products of these anticommutators ensuring closure under both commutators and anticommutators. Following this procedure, the initial gauge group is extended from $SO(1,4)$ to $SO(2,4)\times U(1)$.

\subsection{Fuzzy Gravity}\ 

With the gauge group fixed, we can now begin with the formulation of the wanted gauge theory on the above covariant, noncommutative (fuzzy) background space.

We begin by defining the covariant coordinate:
\begin{equation}\label{CovariantCoordinate}
\mathcal{X}_\mu = X_\mu \otimes \mathbb{1}_4 + A_\mu(X),
\end{equation}
where $A_\mu$ is the gauge connection, which expands on the generators of the gauge group as
\begin{equation}\label{GaugeConnectionFuzzy}
A_\mu = a_\mu \otimes \mathbb{1}_4 + \omega_\mu{}^{ab}\otimes M_{ab} + e_\mu{}^a \otimes P_a + b_\mu{}^a \otimes K_a + \tilde{a}_\mu \otimes D.
\end{equation}
The covariant coordinate then reads
\begin{equation}
\mathcal{X}_\mu = (X_\mu + a_\mu) \otimes \mathbb{1}_4 + \omega_\mu{}^{ab}\otimes M_{ab} + e_\mu{}^a \otimes P_a + b_\mu{}^a \otimes K_a + \tilde{a}_\mu \otimes D.
\end{equation}
Followingly, the covariant noncommutative field strength tensor is defined as \cite{Madore_1992, Manolakos_paper1}:
\begin{equation}
\hat{F}_{\mu\nu} \equiv [\mathcal{X}_\mu, \mathcal{X}_\nu] - \kappa^2 \hat{\Theta}_{\mu\nu},
\end{equation}
where $\hat{\Theta}_{\mu\nu} \equiv \Theta_{\mu\nu} + \mathcal{B}_{\mu\nu}$, and $\mathcal{B}_{\mu\nu}$ is a 2-form promoting $\Theta$ to its covariant form. As an element of the gauge algebra, $\hat{F}_{\mu\nu}$ can be expanded as
\begin{equation}
\hat{F}_{\mu\nu} = R_{\mu\nu} \otimes \mathbb{1}_4 + \frac{1}{2} R_{\mu\nu}{}^{ab}\otimes M_{ab} + \tilde{R}_{\mu\nu}{}^a \otimes P_a + R_{\mu\nu}{}^a \otimes K_a + \tilde{R}_{\mu\nu}\otimes D.
\end{equation}

An SSB follows, which proceeds analogously to the CG case. A scalar field $\Phi(X)$ in the second-rank antisymmetric rep of $SO(2,4)$ is introduced in the action and fixed in a gauge yielding the Lorentz group \cite{Manolakos_paper1, Manolakos_paper2, Roumelioti:2024lvn}. The scalar is also taken to be charged under $U(1)$ in order to break this symmetry. The resulting action reads:
\begin{equation}
\mathcal{S} = \operatorname{Trtr} \Big[ \lambda \Phi(X) \varepsilon^{\mu\nu\rho\sigma} \hat{F}_{\mu\nu} \hat{F}_{\rho\sigma} + \eta \left(\Phi(X)^2 - \lambda^{-2} \mathbb{1}_N \otimes \mathbb{1}_4 \right) \Big],
\end{equation}
where $\eta$ is a Lagrange multiplier and $\lambda$ is a dimensionful parameter. After the SSB, the residual gauge symmetry of the broken action is $SO(1,3)$. As shown in \cite{Manolakos_paper2}, the commutative limit of this action reduces to the Palatini action, which is equivalent to EG with a cosmological constant term.

\section{Unification of Conformal and Fuzzy Gravities with Internal Interactions, Fermions, and Symmetry Breakings}

In \cite{Roumelioti:2024lvn}, it was proposed that CG can be unified with internal interactions in a framework that naturally leads to an $SO(10)$ Grand Unified Theory (GUT), by using $SO(2,16)$ as the single unification gauge group. This choice for the unification gauge group is motivated from the facts that:
\begin{itemize}
    \item It should be possible to reach both the $SO(2,4)$ and $SO(10)$ gauge groups through SSBs, starting from the initial unification gauge group and
    \item In order to have a chiral theory, we need a group of the form $SO(4n+2)$.
\end{itemize}
Given the above requirements, it becomes evident that the smallest unification group which satisfies them is the $SO(2,16)$. As highlighted in the Introduction, the key idea relies on the fact that the tangent space dimension need not coincide with that of the underlying manifold \cite{Weinberg:1984ke, roumelioti2407, Percacci:1984ai, Percacci_1991, Nesti_2008, Nesti_2010, Krasnov:2017epi, Chamseddine2010, Chamseddine2016, noncomtomos, Konitopoulos:2023wst}.  

In what follows, for reasons of simplicity, we work with the Euclidean signature (the implications of the non-compact case is discussed in detail in \cite{Roumelioti:2024lvn}). Starting from $SO(18)\sim SO(2,16)$ with fermions in its spinor rep, $\mathbf{256}$, the SSB leads to its maximal subgroup $SO(6)\times SO(12)$ \cite{Roumelioti:2024lvn}. The relevant branching rules are:
\begin{equation}\label{so18}
\begin{aligned}
SO(18) & \supset SO(6) \times SO(12) \\
\mathbf{256} & = (\mathbf{4}, \overline{\mathbf{32}}) + (\overline{\mathbf{4}}, \mathbf{32}) 
  && \text{(spinor)} \\
{\mathbf{153}} & =(\mathbf{15}, \mathbf{1}) + (\mathbf{6}, \mathbf{12}) + (\mathbf{1}, \mathbf{66}) & & \text {(adjoint)} \\
\mathbf{170} & = (\mathbf{1}, \mathbf{1}) + (\mathbf{6}, \mathbf{12}) + (\mathbf{20}', \mathbf{1}) + (\mathbf{1}, \mathbf{77}) 
  && \text{(2nd rank symmetric)}
\end{aligned}
\end{equation}

The breaking of $SO(18)$ to $SO(6) \times SO(12)$ is triggered by assigning a vev to the $(\mathbf{1},\mathbf{1})$ component of a scalar in the $\mathbf{170}$ rep with fermions in the $\mathbf{256}$ spinor rep of $SO(18)$.

In order to further break $SO(12)$ down to $SO(10) \times U(1)$ or $SO(10) \times U(1)_{\text{global}}$, we can employ scalar fields from the $\mathbf{66}$ representation (contained within the adjoint $\mathbf{153}$ of $SO(18)$) or the $\mathbf{77}$ representation (contained within the 2nd rank symmetric tensor representation $\mathbf{170}$ of $SO(18)$), respectively, given the following branching rules:
\begin{equation}
\begin{aligned}
SO(12) & \supset SO(10) \times \left[U(1)\right]\\
\mathbf{66} & =(\mathbf{1})(0)+( \mathbf{10})(2)+( \mathbf{10})(-2)+( \mathbf{45})(0) \\
\mathbf{77} & =(\mathbf{1})(4)+( \mathbf{1})(0)+( \mathbf{1})(-4)+( \mathbf{10})(2)+( \mathbf{10})(-2)+( \mathbf{54})(0) \\
\end{aligned}
\end{equation}
where the $\left[{U}(1)\right]$ above is there to take into account that the $U(1)$ either remains as a gauge symmetry, or it is broken leaving a $U(1)$ as a residual global symmetry. Given the above branching rules, a VEV to the $<$$(\mathbf{1})(0)$$>$ component of the $\mathbf{66}$ representation leads to the gauge group $SO(10) \times U(1)$ after the SSB, while a VEV to the $<$$(\mathbf{1})(4)$$>$ component of the $\mathbf{77}$ representation results in $SO(10) \times U(1)_{\text{global}}$.

Similarly, we can further break $SU(4)\sim SO(6)$ down to $SO(4) \sim SU(2) \times SU(2)$ in two stages. In the first stage it breaks to $SO(2,3) \sim SO(5)$ and then, in the second stage, to $SO(4)$ according to the following branching rules \cite{Slansky:1981yr}:
\begin{equation}
\begin{aligned}
SU(4) & \supset SO(5)\\
\textbf{4} & =\textbf{4}\\
\textbf{6} & =\textbf{1}+\textbf{5}\\
\end{aligned}
\end{equation}
As an initial step, by assigning a VEV to the $<$$\mathbf{1}$$>$ component of a scalar in the $\mathbf{6}$ representation of $SU(4)$, the latter breaks down to $SO(5)$. Then, according to the branching rules \eqref{SO5toSU2SU2}, 
by giving a VEV to the $<$$\mathbf{1},\mathbf{1}$$>$ component of a scalar in the $\mathbf{5}$ representation of $SO(5)$, we finally obtain the Lorentz group $SU(2) \times SU(2) \sim SO(4) \sim SO(1,3)$. Additionally, it is notable that in this scenario, the $\mathbf{4}$ representation decomposes under $SU(2) \times SU(2) \sim SO(1,3)$ into the appropriate representations to describe two Weyl spinors.

One can also follow an alternative route to break $SU(4)$ to $SU(2) \times SU(2)$, just like in the CG case. Specifically, in order to break the $SU(4)$ gauge group to $SU(2) \times SU(2)$, we can use scalars in the adjoint $\mathbf{15}$ representation of $SU(4)$, which is contained in the adjoint $\mathbf{153}$ representation of $SO(18)$. In this case, we have:
\begin{equation}
\begin{aligned}
SU(4) \supset & SU(2) \times SU(2) \times U(1)\\
\textbf{4} = & (\textbf{2},\textbf{1})(1)+(\textbf{1},\textbf{2})(-1)\\
\textbf{15}  = & (\textbf{1},\textbf{1})(0)+(\textbf{2},\textbf{2})(2)+(\textbf{2},\textbf{2})(-2)\\
&+(\textbf{3},\textbf{1})(0)+(\textbf{1},\textbf{3})(0),
\end{aligned}
\end{equation}
from where, by assigning a VEV to the $<$$\mathbf{1},\mathbf{1}$$>$ direction of the adjoint  representation $\mathbf{15}$, we obtain the known result \cite{Li:1973mq} that $SU(4)$ spontaneously breaks to $SU(2) \times SU(2) \times U(1)$. The method for eliminating the corresponding $U(1)$ gauge boson and retaining only $SU(2) \times SU(2)$ is the same as in the CG case. Again, note that the $\mathbf{4}$ representation decomposes into the appropriate representations of $SU(2) \times SU(2) \sim SO(1,3)$ suitable for describing two Weyl spinors.

Having established the analysis of various symmetry breakings using branching rules under maximal subgroups, starting from the group $SO(18)$, one can correspondingly consider instead the isomorphic algebras of the various groups. Specifically, instead of $SO(18)$, one can consider the isomorphic algebra of the non-compact groups $SO(2,16) \sim SO(18)$, and similarly $SO(2,4) \sim SO(6) \sim SU(4)$.

Consequently, after all the breakings, we obtain:
\begin{equation}
\begin{gathered}
    SU(2)\times SU(2) \times SO(10) \times [U(1)]\\
    \{(\textbf{2},\textbf{1})+(\textbf{1},\textbf{2})\}\{\textbf{16}(-1)+\overline{\textbf{16}}(1)\}+\{(\textbf{2},\textbf{1})+(\textbf{1},\textbf{2})\}\{\overline{\textbf{16}}(1)+\textbf{16}(-1)\}\\
    =2\times\textbf{16}_L(-1)+2\times\overline{\textbf{16}}_L(1)+2\times\textbf{16}_R(-1)+2\times\overline{\textbf{16}}_R(1), 
\end{gathered}
\end{equation}
from where, given that $\overline{\textbf{16}}_R(1)=\textbf{16}_L(-1)$ and $\overline{\textbf{16}}_L(1)=\textbf{16}_R(-1)$, and by choosing to keep only the $\textbf{-1}$ eigenvalue of $\gamma^5$, we obtain
\begin{equation}
    4\times \textbf{16}_L(-1)\, .
\end{equation}
Therefore, this construction yields a natural prediction of four fermion families, arising from the underlying group-theoretic structure. The flavour separation is left as an open problem for future work.

For the Fuzzy Gravity case, in ref \cite{roumelioti2407} it is noted that unifying FG with internal interactions requires fermions to:
\begin{itemize}
    \item be chiral to remain light at low energies, and 
    \item appear in matrix reps consistent with the matrix model construction of FG.
\end{itemize}
This is achieved by starting with the $SO(6)\times SO(12)$ gauge theory and fermions in $(\mathbf{4}, \overline{\mathbf{32}})+(\overline{\mathbf{4}}, \mathbf{32})$, thus satisfying both criteria. Additionally, the gauge-theoretic formulation of gravity in FG requires gauging $SO(2,4) \times U(1) \sim SO(6)\times U(1)$, leading to a low-energy structure closely analogous to the CG case.

\section{Conclusions}

In \cite{Roumelioti:2024lvn}, a potentially realistic framework was developed in which gravity and internal interactions in four dimensions are unified by gauging an enlarged tangent Lorentz group. This approach relies on the key observation that the tangent space dimension can exceed that of the underlying manifold. By constructing CG as a gauge theory of $SO(2,4)$ and implementing spontaneous symmetry breaking, both Einstein Gravity and Weyl Gravity emerge as possible low-energy limits. 

Extending this framework to include internal interactions via $SO(10)$ GUTs was achieved using the higher-dimensional tangent group $SO(2,16)$, with fermions subject to the Weyl condition. A parallel construction for Fuzzy Gravity \cite{roumelioti2407} starts from $SO(2,4)\times SO(12)$ with fermions in $(\mathbf{4}, \overline{\mathbf{32}})+(\overline{\mathbf{4}}, \mathbf{32})$, leading to a unified, gauge-theoretic description of fuzzy gravity and internal interactions.

The low energy limit of the above construction has been studied in \cite{Patellis-Z-24}, by the employment of a 1-loop analysis. Four channels of breaking $SO(10)$ down to the SM have been explored, providing estimates for all the breaking scales from the Planck scale down to the EG scale.

\paragraph{Data Availability Statement:} No data are associated with this manuscript.

\printbibliography

\end{document}